\documentstyle[aps,prl,twocolumn]{revtex}
\begin{document}
\title{High Temperature Expansion Study of the
 Nishimori multicritical Point in Two and Four Dimensions}
\author{Rajiv R. P. Singh}
\address{Department of Physics, University of California,
Davis, CA 95616}
\author{Joan Adler}
\address{Department of Physics, Technion-Israel Institute of  
Technology, 32000, Haifa, Israel}
\date{\today}
\maketitle

\begin{abstract}
We study the two and four dimensional Nishimori multicritical point
via high temperature expansions for
the $\pm J$ distribution, random-bond, Ising model.
In $2d$ we estimate the
the critical exponents along the Nishimori line
to be $\gamma=2.37\pm 0.05$,
$\nu=1.32\pm 0.08$. These, and earlier $3d$ estimates  $\gamma
=1.80\pm 0.15$, $\nu=0.85\pm 0.08$ are remarkably close to
the critical exponents for percolation, which are known to
be $\gamma=43/18$, $\nu=4/3$ in $d=2$ and $\gamma=1.805\pm0.02$
and $\nu=0.875\pm 0.008$ in $d=3$. However, the estimated 
$4d$ Nishimori exponents 
$\gamma=1.80\pm 0.15$, $\nu=1.0\pm 0.1$,
are quite distinct from the $4d$ percolation results
$\gamma=1.435\pm 0.015$, $\nu=0.678\pm 0.05$.
\end{abstract}
\pacs{PACS:}

\narrowtext

In recent years there has been much interest in the
study of critical phenomena in quenched-random, 
two-dimensional, thermodynamic systems.
However, with the exception of percolation, for which
various critical parameters are known exactly, other random
fixed points are not fully understood. Such random 
critical phenomena are of interest from a theoretical
point of view, and also from an
experimental point of view. Among notable experimental systems
showing two-dimensional random critical phenomena are
pleateau transitions in quantum hall systems \cite{qhall} 
and Bose-glass transitions in dirty 
superfluids and superconductors \cite{boseglass}.

Perhaps the simplest theoretical model with quenched randomness is
the random-bond Ising model. In $d=2$ a lot is known  about
weak randomness \cite{weak}. The case where randomness has the most
dramatic influence on thermodynamic properties
is that of a symmetric distribution of bonds,
that is one where there are roughly equal
tendencies for ferromagnetic and
antiferromagnetic ordering. In this case the system may only have
a long-ranged spin-glass phase at low temperatures.
There is considerable numerical evidence that in $d=2$
there is no finite temperature spin-glass phase
\cite{symmetric}. The Nishimori manifold 
separates the region in parameter space where
ferromagnetic ( or antiferromagnetic ) correlations are stronger
from those where spin-glass correlations dominate. In certain
random-bond Ising models many exact results can be obtained along
this special manifold \cite{nishi}. A 
Nishimori multicritical point can exist even
in the absence of a finite temperature spin-glass transition
and has been studied by
renormalization group \cite{rg} and 
various numerical methods \cite{numerics}. By now
the existence of the critical point and its location are
reasonably well established \cite{numerics}, 
although to our knowledge no
reliable estimates of the critical exponents exist.

Here we study this model by high temperature expansions,
estimating the critical point and the various critical exponents.
Our estimates for the critical temperatures are consistent 
with previous ones.
Our interesting result is that the critical exponents $\gamma$,
and $\nu$ are remarkably close to that of percolation. 
This turns out to be the case in three dimensions also. To
see if such a trend continues with dimensionality we study
the four dimensional Nishimori multicritical point. 
There the critical exponents are 
clearly distinct from percolation. 
Towards the end of this paper we speculate on the relevance of
the percolation fixed point to the 
present problem.

We consider the Hamiltonian
\begin{equation}
{\cal H}=\sum_{<i,j>} J_{i,j} S_i S_j 
\end{equation}
with the $J_{ij}$ independent quenched random variable with distribution
\begin{equation}
P(J_{ij})=p\delta(J_{ij}-J)+(1-p)\delta(J_{ij}+J)
\end{equation}
The Nishimori line in the parameter space of temperature $T$ and
ferromagnetic bond concentration $p$ is given by
\begin{equation}
v=2p-1 
\end{equation}
with $v=\tanh{J/kT}$.
It is most convenient to directly develop the expansions
along the Nishimori line \cite{d=3}. In this case the expansion
variable is $w=v^2$.  We define susceptibilities
$\chi_{m,n}$ by the relations
\begin{equation}
\chi_{m,n}={1\over N}\sum_{i,j} [<S_i S_j>^m]^n
\end{equation}
Here angular brackets represent thermal averaging and the square brackets
an averaging with respect to the bond distribution.
We carry out expansions to 19th order in $2d$ and 15th order in $4d$
for $\chi_{2,1}$ and $\chi_{2,2}$ using the star-graph
method \cite{star}. We note that along the
Nishimori line the ferromagnetic susceptibility, $\chi_{1,1}$
exactly equals the spin-glass susceptibility $\chi_{2,1}$. In order to
study the crossover exponents we also calculate the series for
$\chi_v=v\partial\chi_{2,1}/\partial v$.The 
expansion coefficients are given in table 1.

We shall assume that the quantities $\chi\equiv\chi_{2,1}$, 
$\chi^\prime\equiv\chi_{2,2}$
and $\chi_v$, become singular at a critical point $w_c$
with exponents $\gamma$, $\gamma^\prime$ and $\gamma^{\prime\prime}$
respectively.
Assuming standard scaling, the exponents for the divergence
of the different series can be related to the critical exponents
$\nu$, $\eta$ and $\phi$ as \cite{d=3}:
$$\gamma=(2-\eta)\nu,\qquad \gamma^\prime=(d-4-2\eta)\nu,
\qquad \gamma^{\prime\prime}=\gamma+\phi$$
Here $d$ is the dimensionality of the system.

We analyze the series based on the expectation that near the
critical point the susceptibilities have the form
$$\chi
\propto (w_c-w)^{-\gamma}(1+a(w_c-w)^{\Delta_1}+b(w-w_c)+\cdots).$$
We estimate the location of the critical point, $w_c$,
the  dominant  exponent $\gamma$, and the correction-to-scaling
exponent $\Delta_1$.
The value of $\Delta_1$  may be biassed by the presence
of even higher order correction terms, however fitting to this form should 
ensure reliable evaluation of the critical point and 
the dominant exponent.
Our analysis is
carried out with no  prior assumptions regarding exponent values,
and  was done
by one of us without knowing prior literature values for the
critical parameters
or scaling relations between the exponents.
We have studied  the series   with two methods,
 commonly known as M1 and M2
\cite{methods}.
They are based on suitable transformations of the series,
 and Pad\'e
approximants for the transformed series.

\noindent {\bf M1:} In  this  method of analysis we study the
logarithmic derivative of
$$
B(w)=hH(w)-(w_c-w){{{\rm d}H(w)}\over{{\rm d}w}}, $$
The dominant singularity is a pole at $w=w_c$ with a residue
$(  h-1)$.  

 We 
implement method M1 as follows: for a given value of $w_c$ we obtain
$\Delta_1$ versus input $h$ for many central and high
 Pad\'e approximants, and we choose the
triplet $w_c, h, \Delta_1$ where all Pad\'es yield as nearly as possible
identical values of $h$.

\noindent {\bf M2:} In the second method
 we first transform the series in $w$ into a
series in the variable $y$, where
$$y=1-(1-w/w_c)^{\Delta_1},  $$
and then take Pad\'e approximants to
$$G(y)  =\Delta_1   (y - 1)  {{\rm d}\over{{ \rm d}y}} \ln (H(w))
, $$
which should converge to $-h$.  Here we plot graphs of $h$ versus
the input $\Delta_1$ for different values of $w_c$ and again choose the
triplet $w_c, h, \Delta_1$,
 where all Pad\'es converge to the
same point.  

We found good convergence for all three $d=2$ series, with M2 graphs
being better converged everywhere, and M1 giving consistent results.
The $\chi$ and $\chi^{\prime}$ series behaved better than 
the $\chi_v$ series.
For brevity, we only show some representative plots
from which we have deduced our estimates of critical parameters.
In  Figure 1 we present two three dimensional graphs from the
 M2 analysis for the $\chi$ series.
In Figure 1(a) we show the three dimensional version on a fairly
 coarse temperature 
scale, in  Figure 1(b) we show a finer scale. 
On the coarse scale we can see that at the trial $w_c$ values of
0.525 and 0.550, convergence to a region of clear intersections is
much poorer than at 0.575. Similarly, although the different approximants
come together at the background plane with $w_c=0.625$, this type 
of almost flat graph  with the asymptotic convergence at very high $\Delta_1$
values is indicative of behaviour that does not give correct critical 
behaviour in test systems or exactly solved models. (It is often seen near
trial critical points that give
exponent values that violate hyperscaling.)
 The fine scale shown in the enlargement
in Figure 1(b)  shows us a set of graphs which mostly satisfy the 
considerations required of 
an intersection region, with the best of all being the central plane.
 Thus from the M2 analysis, the best 
$w_c$ estimate is $0.596\pm0.008$.
This implies $p_c=0.886\pm0.003, T_c/J=0.975\pm 0.006$, which
are consistent with previous estimates \cite{numerics}.
    In  Figure 2(a) the central slice
at $w_c=0.596$ is shown. 
 From this  we conclude  an exponent estimate of $\gamma=2.37\pm0.05$.
The M1 analysis is consistent with these values. 
In the
 M2 analysis for the $\chi^{\prime}$ series,
 convergence was again optimal at $w_c=0.596$.
The exponent, $\gamma^{\prime}$, is deduced to be $2.11\pm0.07$.
Via scaling this gives $\nu=1.32\pm0.08$.  
We analyze the $\chi_v$ series in two ways, first by considering
$\chi_v/w$ and second by studying the series for 
$d\chi_v/dw$.
From these analyses 
we conclude $\gamma+\phi=3.0\pm0.3$.

In four dimensions, the M2 analysis of the $\chi$ series gives 
best  convergence
at $w_c=0.1764$, with
$\gamma= 1.9$.
From M1 a slightly lower $w_c=0.176$ value
 and a correspondingly lower
$\gamma=1.8$ seems optimal.
For the  $\chi^{\prime}$ series optimal convergence is at $w_c=0.176$ with 
the central values of the exponent $\gamma^{\prime}$ of $0.41$ from M1
and $0.40$ from M2. 
Since three of the four analysis support the lower
value of $w_c=0.176$ we use this for our final estimates.
We quote overall values of
$w_c=0.176\pm0.001$ and $\gamma=1.80\pm0.15$,  $\gamma^{\prime}= 0.40\pm0.03$
By scaling this gives us $\nu=1.0\pm0.1$.

As stated in the introduction, these estimated critical parameters are
remarkably close to percolation \cite{percolation2} in $d=2$.
The numbers for percolation are
$\gamma=43/18$ and $\nu=4/3$. Furthermore, in $d=3$ the 
Nishimori exponents were found to be $\gamma=1.80\pm 0.15$ and $
\nu=0.85\pm 0.08$ \cite{d=3}. 
These numbers are also confirmed by our present analysis.
The $d=3$ percolation exponents are
$\gamma=1.805\pm0.02$
 and  $\nu=0.875\pm0.008$ \cite{percolation2}.  Thus,
in $d=3$ also the Nishimori exponents are very close to percolation.
However, in $d=4$
the percolation exponents are $\gamma=1.435\pm 0.015$ and
$\nu=0.678\pm 0.05$, which are clearly distinct from those
found here. We note also that the exponent $\nu$ along
the Nishimori line appears
to be non-monotonic as a function of dimensionality.

One can conclude that in $d=2$ and $3$ the
Nishimori critical behavior is consistent with the
universality class for percolation. 
However, this is not so in $d=4$. 
The possibility that the closeness of the Nishimori exponents
to percolation in $d=2$ and $3$ is purely accidental
cannot be ruled out. However, the following considerations
suggest a possible connection. 
If we consider a bond distribution
of the form:
$$P(J_{ij})=p\delta(J_{ij}-J)+(1-p)\delta(J_{ij}+J)+c\delta(J_{ij})$$
That is, the bonds are allowed to take values $\pm J$ as
well as zero, then in the generalized parameter space of $p$, $c$
and temperature $T$, the Nishimori manifold is a 
two-dimensional plane. For 
$p=0$ ( or unity),
this plane reduces to the $T=0$ dilution axis and thus contains the
percolation fixed point \cite{hiroguchi}.
However, it is generally believed 
that T is always an unstable direction for percolation
and, hence, finite temperature Nishimori criticality should have
different exponents. 

Secondly, Nishimori has argued \cite{nishi2} that the
spin-glass ferromagnetic transition is a 
{\it geometry-induced} phase transition ( as opposed to a
thermal transition ), which is also true for percolation.
However, if this led to the identification of the Nishimori
fixed point with percolation it should be true independent
of dimensions. However, our results in $d=4$ contradict this.
Furthermore, the epsillon expansions (around $d=6$)
for the exponents at the Nishimori multicritical point are
different from percolation \cite{rg}.

Finally, a very different way in which this model is of 
significant interest,
is through the mapping between the $2d$ Ising model and free
fermions in $1+1$-dimension, and the connection between the
latter and the pleateau transitions in the quantum hall effect.
The Chalker model for the
pleateau transitions in the quantum hall effect
\cite{chalker,ludwig} can thus
be mapped onto random-bond Ising models. However, the $\pm J$
model studied here
does not have the correct symmetries for the quantum-hall
problem \cite{mpaf}. It is well known that percolation occurs in one
limit of the quantum-hall systems, when the disorder potential
is slowly varying in space \cite{trugman}. However, numerical
studies of the Chalker models lead to exponents clearly
different from percolation \cite{chalker}. 

Thus, there is no compelling 
theoretical reason why the Nishimori multicritical point should
be in the universality class of percolation.
One possible explanation for our findings 
could be that in low-dimensions,
where the multicritical point occurs at very low temperatures,
there are crossover effects which produce effective exponents
close to percolation.
These issues deserve further attention.

ACKNOWLEDGEMENTS

We would like to thank Matthew Fisher and Michael Fisher
for many valuable
discussions. RRPS is supported in part by
NSF grant number DMR-93--18537.
Part of this project was completed while 
JA visited UC Davis with support
from the US -Israel Binational Science Foundation.

\begin{figure}
\caption{Plots from the M2 analysis of the $d=2$
$\chi$ series (a) on a coarse scale and (b) on a fine scale.}
\end{figure}

\begin{figure}
\caption{The central slice for the M2 analysis in Figure 1
with $w_c=0.596$. }
\end{figure}

\begin{table}
\caption{ Expansion coefficients for the susceptibilities
in $d=2$ ( See Eq. 4)}
\begin{tabular}{rrrr}
$n$&$\chi_{2,1}$&$\chi_{2,2}$&$\chi_v$\\
\hline
0& 1 & 1 &0 \\
1& 4 & 0 &8 \\
2& 12 & 4&48 \\
3& 36 & 0 &216\\
4& 76 & 36 &512 \\
5& 196 & -32 &1640\\
6& 316 & 236 &1856\\
7& 884 & -464 &9208\\
8& 780 & 1988 &-5824\\
9& 3684 & -5072 &57576\\
10& 396 & 17076 &-109264\\
11& 22740 &-50432 &680376\\
12& -22596 & 164108 &-1547376\\
13& 188420 & -500496 &8163624\\
14& -331108 & 1604572 &-21618432\\
15& 1517396 & -5042160 &84085560\\
16& -4509268 & 16221028 &-311253632\\
17& 15654148 & -52336864 &1071437960\\
18& -56714548 & 170687620 & -4263416944\\
19& 183041524 & -561493296& 14787979576\\
\hline 
\end{tabular}
\label{d=2}
\end{table}

\begin{table}
\caption{ Expansion coefficients for the susceptibilities
in $d=4$ ( See Eq. 4)}
\begin{tabular}{rrrr}
\hline
$n$&$\chi_{2,1}$&$\chi_{2,2}$&$\chi_v$\\
\hline
0& 1 & 1 &0\\
1& 8 & 0 &16\\
2& 56 & 8 &224\\
3& 392 & 0 &2352\\
4& 2552 & 200 &19840\\
5& 16904 &-192 &162512\\
6& 105944 &6584 &1179840\\
7& 679784 &-12384 &8736880\\
8& 4158200 &234824 &58846592\\
9& 26120392 &-649056 &412368720\\
10& 157020984 &8748712 &2651405536\\
11& 974362408 &-32109952 &18054488432\\
12& 5783009304 &342786296 &112459651552\\
13& 35661616648 &-1523180000 &755621878608\\
14& 209506120728 &14008147224 &4590427798720\\
15& 1289118273320 &-70814307872 &30721400183024\\
\hline 
\end{tabular}
\label{d=4}
\end{table}

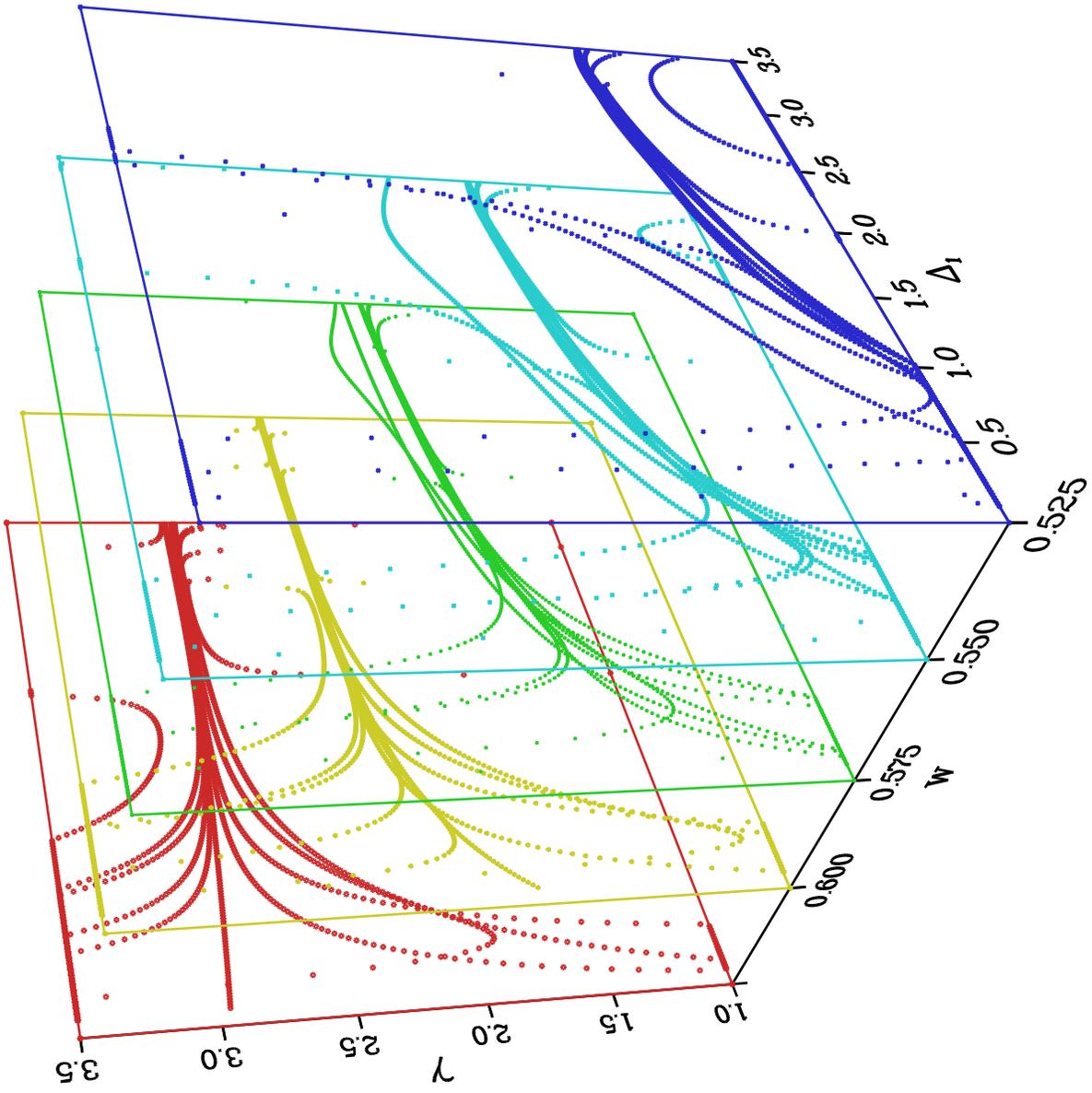
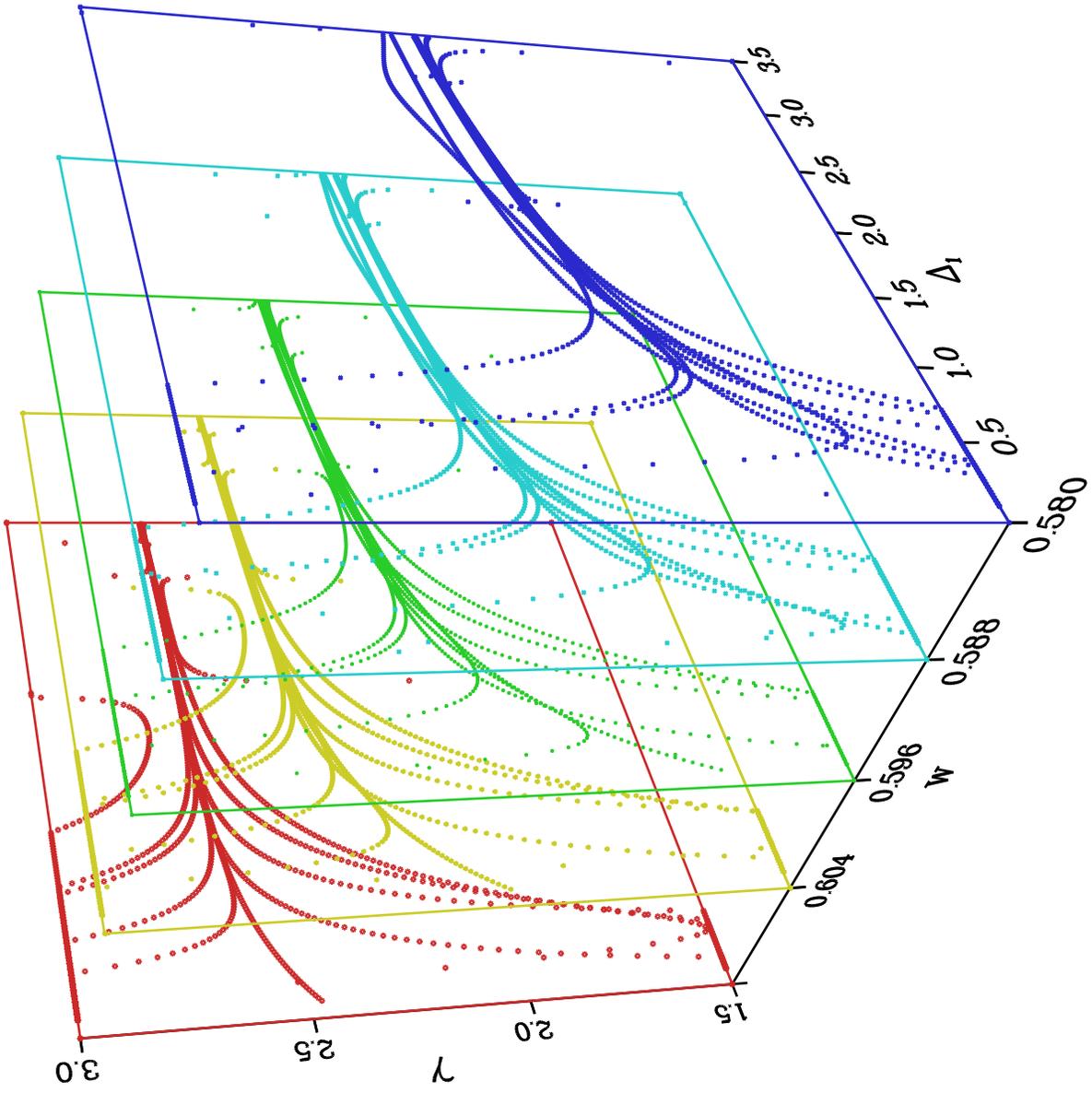
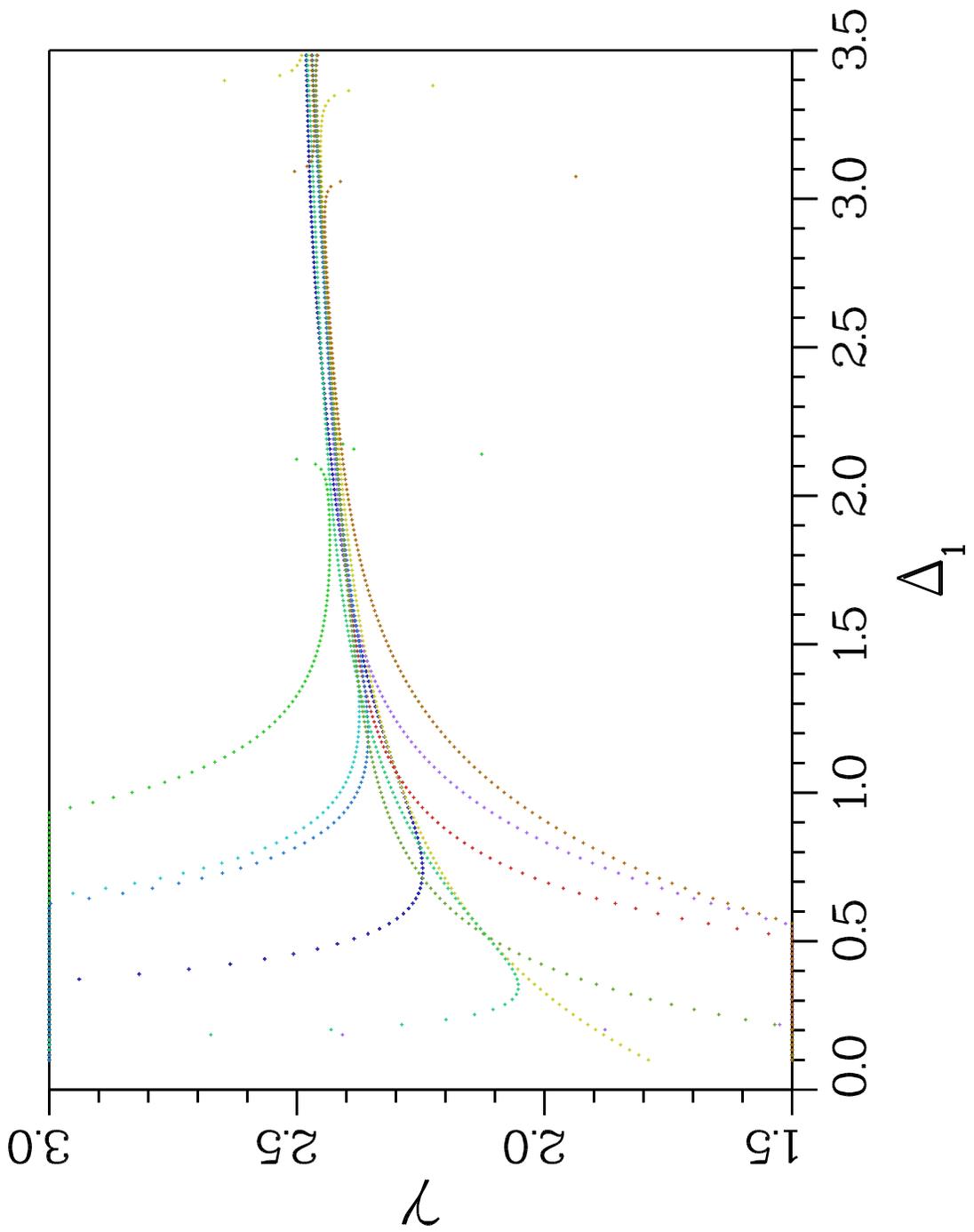
\end{document}